# Implications of particle acceleration in active galactic nuclei for cosmic rays and high energy neutrino astronomy[1]

A.P. Szabo[2] and R.J. Protheroe

*Department of Physics and Mathematical Physics, University of Adelaide, SA 5005, Australia*

We consider the production of high energy neutrinos and cosmic rays in radio-quiet active galactic nuclei (AGN) or in the central regions of radio-loud AGN. We use a model in which acceleration of protons takes place at a shock in an accretion flow onto a supermassive black hole, and follow the cascade that results from interactions of the accelerated protons in the AGN environment. We use our results to estimate the diffuse high energy neutrino intensity and cosmic ray intensity due to AGN. We discuss our results in the context of high energy neutrino telescopes under construction, and measurements of the cosmic ray composition in the region of the "knee" in the energy spectrum at $\sim 10^7$ GeV.

## 1 Introduction

Active galactic nuclei have long been thought to be sites of energetic particle production [1, 2]. For example, observations of the X-ray emissions from AGN [3, 4, 5] indicate that there is probably a non-thermal distribution of electrons present with energies up to at least 1 MeV (see e.g. [6]). X-ray variability on time scales as short as $\sim 100$ s (e.g. [7]) imply that the emission comes from a compact region within the AGN. Recently, the EGRET experiment on the Compton Gamma Ray Observatory has detected $\gamma$-ray emission from at least 25 AGN (see the 1st EGRET source catalog [8] and references therein), with one object, Mrk 421, also being detected at TeV energies by the ground based Whipple telescope [9]. However, all of the AGN detected thus far by EGRET have been blazars, and so it seems likely that the $\gamma$-ray emission from these objects is associated with jets. Blazars, however, form a small fraction of all AGN, and we wish to consider the neutrino and cosmic ray emission from the much more numerous 'ordinary' AGN (e.g. Seyferts and radio-quiet AGN), or from the central engines of blazars as distinct from the

---

jets.

AGN were first suggested as possible sources of high energy neutrinos in the 1970s. Berezinsky and Smirnov [10] and Berezinskiĭ and Zatsepin [11] calculated the flux of neutrinos resulting from the decay of pions produced in *pp* collisions of cosmic rays escaping from AGN. They argued that AGN would be sources of cosmic rays with energies up to $\sim 10^{10}$ GeV and that these high energy particles would interact with the matter in the galaxy or the intergalactic medium or with the cosmic microwave background. Neutral pions are produced in these interactions, and this enabled them to derive an upper limit for the neutrino flux from AGN based on the observed X-ray and $\gamma$-ray fluxes. However, AGN probably contain sufficient matter to stop most protons before they diffuse out of the galaxy. This led Eichler [12] to suggest that AGN may be extremely luminous high energy neutrino sources, with the power radiated as neutrinos being comparable to the total power in all other energy bands. Eichler also suggested that the neutrino background due to unresolved AGN may be detectable with the neutrino telescopes such as DUMAND being discussed at the time.

The importance of $p\gamma$ interactions *within* AGN was first noted by Sikora *et al.* [13]. X-ray observations, in particular, indicate that AGN are very compact and hence the radiation density within the nucleus of the galaxy is likely to be large. In fact, at high energies, Sikora *et al.* found that $p\gamma$ interactions (both pion photoproduction and pair production) become the dominant loss processes and may determine the highest particle energy achievable in AGN.

Neutrons are by-products of both *pp* and $p\gamma$ interactions. Unlike protons, neutrons are not confined magnetically within the galactic nucleus and may provide an effective way of transporting energy out of the nucleus. This mechanism may also be responsible for the production of an observable flux of $\gamma$-rays [14, 15, 16, 17, 18] and may also result in an AGN component of the observed cosmic rays at $\sim 10^6 - \sim 10^7$ GeV [19].

Recently, Stecker *et al.* [20] calculated the intensity of neutrinos expected from radio-quiet AGN. They found that while individual AGN are unobservable with the proposed neutrino telescopes (e.g. DUMAND, AMANDA, NESTOR, BAIKAL, etc., see [21, 22, 23] for reviews), the diffuse neutrino flux from unresolved AGN should easily be observable. Their result generated much interest amongst the high energy neutrino astrophysics community. Unfortunately, an error in the published X-ray luminosity function used by Stecker *et al.* meant that their original result was in error by a factor of $\sim 17$ [24]. They have recalculated the diffuse neutrino flux [25, 26] using a different luminosity function and the resulting intensity, while down by a factor of $\sim 30$ on the original estimate, is still interesting.

In calculating the neutrino flux from AGN, Stecker *et al.* only considered $p\gamma \to \Delta^+$ on ultraviolet photons. Here we extend this work to include all of the important loss processes. In Section 2 we describe the model we have used for accretion onto a supermassive black hole, and derive the acceleration rate of protons and nuclei assuming first order Fermi acceleration in the central region of AGN. The relevant loss processes are discussed in Section 3, and the loss rates are used along with the acceleration rates to calculate the maximum particle energy obtainable in AGN. The question of neutron escape from the central region is discussed in Section 4. The Monte Carlo program used to calculate the spectra of particles produced as a result of injecting one proton into the accelerator is described in Section 5. The results are applied in Section 6 to calculate the diffuse neutrino background from AGN and to calculate the contribution from AGN to the intensity of high energy cosmic rays. We discuss our results and give our conclusions in Section 7. Preliminary results of our work have been published in a number conference proceedings [27, 24, 28, 29, 30, 31].

## 2 The Model

The AGN model we adopt is that described by Protheroe and Kazanas [32] and developed by Kazanas and Ellison [33]. Our application of this model is described elsewhere [18], but we also include a brief description here because it is central to the development of the present paper.

The basic ingredients of the model are summarized below. A shock at radius $R = x_1 r_S$, where $r_S$ is the Schwarzschild radius, is assumed to develop in an accretion flow onto a supermassive black hole and be supported by the pressure of relativistic particles. We therefore assume that at the shock the magnetic pressure is comparable to that of relativistic particles and the ram pressure of accreting plasma:

$$\rho_1 u_1^2 \simeq U_p/3 \simeq B^2/8\pi, \qquad (1)$$

where $U_p$ is the energy density in relativistic particles, and $\rho_1$ and $u_1 = x_1^{-1/2} c$ are the density and flow velocity just upstream of the shock. We assume that relativistic particles are responsible for the AGN continuum luminosity, $L_C$, and that about half their energy is lost to neutrinos, i.e.

$$L_C \simeq \frac{1}{2} L_p. \qquad (2)$$

The luminosity in relativistic particles is given by

$$L_p \simeq Q \frac{1}{2} \dot{M} u_1^2 \qquad (3)$$

where $Q(x_1) = 1 - 0.1 x_1^{0.31}$ is the efficiency of conversion of bulk kinetic energy of accreting plasma into energetic particles at the shock [33]. In this model, the black hole mass is proportional to luminosity, and from Figure 5 of Kazanas and Ellison [33] we obtain

$$M \simeq 10^{-38} x_1 L_C \; M_\odot \qquad (4)$$

where $L_C$ is in erg s$^{-1}$. The available data [34], after applying a bolometric correction of ×5, appears to be consistent with a flat distribution in $\log x_1$ for $10 < x_1 < 100$.

We are now in a position to specify the matter, magnetic, and radiation environment in which the particles are accelerated and interact. The matter density just upstream of the shock is related to the accretion rate by

$$\dot{M} = 4\pi R^2 \rho u_1. \qquad (5)$$

Hence, we obtain the matter density just upstream of the shock

$$\rho_1 \simeq 1.4 \times 10^{33} Q^{-1} x_1^{-5/2} L_C^{-1} \quad \text{g cm}^{-3}, \qquad (6)$$

and the magnetic field

$$B \simeq 5.5 \times 10^{27} Q^{-1/2} x_1^{-7/4} L_C^{-1/2} \quad \text{gauss}. \qquad (7)$$

The radiation density in the vicinity of the shock is related to the AGN continuum luminosity by

$$U_{\rm rad} \simeq L_C / \pi R^2 c \qquad (8)$$

giving

$$U_{\rm rad} \simeq 1.2 \times 10^{54} x_1^{-4} L_C^{-1} \quad \text{erg cm}^{-3}. \qquad (9)$$

Making the approximation that the diffusion coefficient at the shock is some constant $b$ times the Bohm diffusion coefficient, i.e. $D = b(\frac{1}{3} r_g c)$ where $r_g$ is the gyroradius, we obtain

$$D = 6.0 \times 10^{-12} b Q^{1/2} x_1^{7/4} L_C^{1/2} Z^{-1} (E/\text{GeV}) \quad \text{cm}^2 \text{ s}^{-1}. \qquad (10)$$

For the simplest case of a parallel shock and equal diffusion coefficients upstream and downstream, the acceleration rate is given by (e.g. [35])

$$\frac{\mathrm{d}E}{\mathrm{d}t} \simeq u_1^2 E / 20 D \qquad (11)$$

and we obtain

$$\frac{\mathrm{d}E}{\mathrm{d}t} \simeq 7.5 \times 10^{30} b^{-1} Q^{-1/2} x_1^{-11/4} L_C^{-1/2} \quad \text{GeV s}^{-1}. \qquad (12)$$

## 3 Loss Processes and Maximum Proton Energies in AGN

The maximum energy obtainable during acceleration will be determined by energy losses associated with particle interactions. The accelerated energetic particle population "cools" by interacting with matter, radiation or magnetic fields. The main loss processes for high energy protons in the central regions of AGN are $p\gamma$ collisions with photons of the infrared–X-ray AGN continuum resulting in either $e^\pm$ pair production or pion production, $pp$ collisions with protons of the accreting plasma, diffusive escape from the acceleration region, advection into the black hole, and synchrotron emission. Szabo and Protheroe [27, 28] describe these processes which occur during and after acceleration of protons in a radiation field (see also Mannheim and Biermann [36], Sikora *et al.* [37], Sikora and Begelman [38], and Berezinsky and Gazizov [39, 40]). In order to see which of the loss process have a significant effect on particle production in the central region, we have calculated the loss time scales for all of the relevant energy loss mechanisms. We define the loss rate as the interaction rate for catastrophic processes (i.e. where there is no particle of the initial type in the final state), or the effective energy-loss rate divided by energy for other processes. For proton-photon interactions, the energy loss time scale is inversely proportional to $U_{\rm rad}$, and so we scale the loss rates for all of the processes by $U_{\rm rad}$ to simplify comparison between them.

The loss rates in $p\gamma$ collisions depend on the spectrum of the AGN continuum radiation. We have used two "generic" photon spectra to represent the continuum radiation in the central region of AGN. The first of these, hereafter spectrum (a), is based on the average AGN spectrum obtained by Saunders *et al.* [41], and is similar to the photon spectrum considered by Stecker *et al.* [20]. It consists of a $dn/d\varepsilon \propto \varepsilon^{-1.7}$ power-law component which extends up to $\sim 1$ MeV, consistent with the component of the X-ray background assumed to come from AGN [6], together with a black body component with temperature $T = 5 \times 10^4$ K used to represent the UV/optical bump which is thought to be thermal emission from an accretion disk [42]. The X-rays are known to be produced at small radii due to spectral variability on short time scales [7]. For spectrum (a) the observed infrared emission from AGN is assumed to come primarily from re-radiation by dust at large radii [43]. Hence, the energy density in infrared radiation in the central region will be negligible compared to that in the UV, optical and X-ray bands. We assume that the energy density contained in the thermal and power-law components is equal.

The second generic photon spectrum we consider, hereafter spectrum (b), is based on the assumption that the infrared emission from AGN is non-thermal in nature, and that the power-law component extends from the infrared to X-ray energies. In this case, we assume a $dn/d\varepsilon \propto \varepsilon^{-2}$ power-law which extends from $10^{-2}$ eV, where Edelson and Malkan [44] have observed a turnover in the spectrum, to $\sim 1$ MeV. Below $10^{-2}$ eV we assume that the source becomes synchrotron-self-absorbed giving a $dn/d\varepsilon \propto \varepsilon^{5/2}$. We again model the UV/optical bump with a black body spectrum, however, in this case we have used a temperature of $T = 2.6 \times 10^4$ K [44]. The energy densities in both the power-law and black body components are again assumed to be equal. Both spectra are shown in Figure 1.

In the remainder of Section 3 we describe how the loss rates for high energy particles in the central region are calculated. The loss rates for protons calculated as described below, and divided by $U_{\rm rad}$, are plotted in Figures 2 and 3.

## 3.1 Proton-Proton Losses

The mean interaction length for proton-proton interactions inside the central region is $\lambda_{pp}(E) = 1/n_2 \sigma_{pp}^{\rm inel}(E)$, where $n_2 \simeq 4\rho_1/m_p$ is the number density downstream of the shock and is obtained using Equation 6, and $\sigma_{pp}^{\text{inel}}(E)$ is the inelastic proton-proton cross section for which we use the data in ref. [45]. The mean inelasticity for proton-proton interactions is $\kappa_{pp} \sim 0.5$ (e.g. [46]) and hence the loss rate divided by $U_{\text{rad}}$ is

$$\frac{1}{U_{\text{rad}}E}\frac{dE}{dt} = \frac{\kappa_{pp}c}{U_{\text{rad}}\lambda_{pp}(E)} \simeq 6.4 \times 10^{10} Q^{-1} x_1^{3/2} \left(\frac{\sigma_{pp}(E)}{\text{cm}^2}\right) \qquad \text{GeV}^{-1}\,\text{s}^{-1}\,\text{cm}^3, \quad (13)$$

where $U_{\text{rad}}$ was obtained from Equation 9.

## 3.2 Pion Photoproduction Losses

The total cross section for pion photoproduction, $\sigma_\pi(s)$, is characterized by a resonance region near the threshold, due primarily to the channels $p\gamma \to \Delta^+ \to p\pi^0$ and $p\gamma \to \Delta^+ \to n\pi^+$. For these reactions, the total centre of momentum (CM) frame energy, $\sqrt{s}$, must exceed the sum of the rest mass energies of the final state pion and nucleon. For a proton of energy $E$ interacting with a photon of energy $\varepsilon$,

$$s = 2\varepsilon E(1 - \beta\cos\theta) \quad (14)$$

where $\theta$ is the angle between the directions of the proton and photon, and $\beta c$ is the proton's velocity. The threshold condition is then $s > s_{\min}$ where $s_{\min} \simeq 1.16$ GeV$^2$. Significantly above threshold, the total cross section is dominated by multiple pion final states and becomes approximately constant with value $\sigma_\pi(s) \simeq 116$ $\mu$b for $s > 10$ GeV$^2$. It is likely that the total cross section may rise at high energies ($s > 100$ GeV$^2$) due to jet events [47]. However, in AGN where energetic protons interact with a radiation field, most of the pion photoproduction interactions occur near threshold, and the very high energy behaviour of the total cross section will have no effect on our results.

For proton-photon interactions the mean interaction length, $\lambda$, is given by,

$$[\lambda(E)]^{-1} = \int_{\varepsilon_{\min}}^{\varepsilon_{\max}} \frac{n(\varepsilon)}{8\varepsilon^2 E^2} \int_{s_{\min}}^{s_{\max}} \sigma(s)(s - m_p^2 c^4) ds\, d\varepsilon, \quad (15)$$

where $n(\varepsilon)$ is the differential photon number density, and $\sigma(s)$ is the appropriate total cross section for the process in question. For pion photoproduction: $\lambda = \lambda_\pi$, $\sigma = \sigma_\pi$, $\varepsilon_{\max} = \infty$, $\varepsilon_{\min} \simeq m_\pi c^2(m_\pi c^2 + 2m_p c^2)/4E$, and $s_{\max} \simeq m_p^2 c^4 + 4\varepsilon E$ which corresponds to a head-on collision of a proton of energy $E$ and a photon of energy $\varepsilon$, assuming $E \gg m_p c^2$. The mean loss rate for pion photoproduction is given by,

$$\frac{1}{E}\frac{dE}{dt} = \frac{\kappa_\pi(E)c}{\lambda_\pi(E)}, \quad (16)$$

where $\kappa_\pi(E) = (E - \langle E'\rangle)/E$ is the mean inelasticity for pion photoproduction and $\langle E'\rangle$ is the mean energy of the nucleon after the interaction. We calculate $\kappa_\pi(E)$ using the Monte Carlo method by sampling a large number of pion photoproduction interactions, as described below, and averaging over the fractional energy lost by the proton in each interaction.

The initial conditions for each interaction are obtained as follows. Examination of the integrand in Equation 15 shows that the energy of the soft photon interacting with a proton of energy $E$ is distributed as

$$p(\varepsilon) = \frac{\lambda_\pi(E) n(\varepsilon)}{8\varepsilon^2 E^2} \Phi_\pi(s_{\max}(\varepsilon, E)) \tag{17}$$

in the range $\varepsilon_{\min} \leq \varepsilon \leq \varepsilon_{\max}$ where

$$\Phi_\pi(s_{\max}) = \int_{s_{\min}}^{s_{\max}} \sigma_\pi(s)(s - m_p^2 c^4) ds. \tag{18}$$

Similarly, examination of the integrand in Equation 15 shows that the square of the total CM frame energy is distributed as

$$p(s) = \frac{\sigma_\pi(s)(s - m_p^2 c^4)}{\Phi_\pi(s_{\max})}, \tag{19}$$

in the range $s_{\min} \leq s \leq s_{\max}$. The rejection technique is used to sample $\varepsilon$ and $s$ respectively from the two distributions, and Equation 14 is used to find $\theta$.

To model the individual interactions we have considered the three channels: (i) $p\gamma \to p\pi^0$; (ii) $p\gamma \to n\pi^+$; and (iii) $p\gamma \to$ multiple pion production. The total pion photoproduction cross section is shown in Figure 4 along with the partial cross sections for the three channels. We have used fits to the data given by Genzel, Joos and Pfeil [48] for the differential cross section for channels (i) and (ii). Integrating these cross sections over all angles gives the partial cross sections $\sigma_{\rm i}(s)$ and $\sigma_{\rm ii}(s)$. The total cross section for multiple pion production is then given by, $\sigma_{\rm iii}(s) = \sigma_\pi(s) - \sigma_{\rm i}(s) - \sigma_{\rm ii}(s)$.

We decide which interaction takes place, for a given value of $s$, on the basis of the total cross sections for the three channels. For example, if a uniform random deviate $\xi$ ($0 \leq \xi \leq 1$) is less than $\sigma_{\rm i}(s)/\sigma_\pi(s)$ then interaction (i) is assumed to occur. If interaction (i) or (ii) occurs then we sample an angle for the pion in the CM frame from the appropriate differential cross section using the rejection technique, and then Lorentz transform the final particles' energies to the laboratory frame.

If interaction (iii) occurs then we use the inclusive data of Moffeit *et al.* [49] to sample the energies and momenta of the produced particles in the CM frame. We do this by computing the invariant cross sections in the CM frame for both leading and central region pions. From these cross sections we derive transverse and longitudinal momentum distributions and calculate the mean multiplicity of leading and central region pions, $n_l$ and $n_c$. We decide whether a given pion's momentum should be sampled from the leading or central pion distributions by selecting a random deviate $\xi$ and testing the condition $\xi < n_l/(n_l + n_c)$. Pions are sampled in this fashion until the energy which remains for particle production is less than $\sqrt{s}/2$, in which case the remaining energy is given to the leading nucleon. Leading pions are produced in the ratio $\pi^+:\pi^-:\pi^0 = 1:1:0$, by analogy with the $\rho^0$ meson decay, whereas central region pions are assumed to be produced with equal numbers of all three charge states. The final-state nucleon is produced on average in the ratio $p{:}n = 3{:}2$ based on counting the various final states the available quarks can produce. Our results are not too sensitive to these assumptions and, when tested, our sampling routine reproduced the 9.3 GeV data of Moffeit et al. [49] to within 15%. In any case, improving the accuracy for these multi-pion events will have little effect on the results as most of the interactions take place near threshold where the accurate fits to the exclusive data in the $\Delta^+$ resonance region are used.

## 3.3 Pair Production Losses

The threshold for this process is $s_{\min} \simeq 0.882$ GeV$^2$. Above threshold, the total cross section for pair production in the field of a proton is approximately 100 times the total cross section for pion photoproduction (see Figure 4). The mean interaction length is obtained from Equation 15 with $\varepsilon_{\min} \simeq m_e c^2 (m_e c^2 + m_p c^2)/E$, $\varepsilon_{\max}$ and $s_{\max}$ being the same as for pion photoproduction, and for the total cross section, $\sigma_e(s)$, we have used the parameterization of Maximon [50].

To model pair production interactions, we first sample $\varepsilon$ and $s$ in a manner directly analogous to that described above for pion photoproduction. We use the formulae for the differential cross section given in the compilation of Motz, Olsen and Koch [51] to model the interaction in the proton's rest frame. The energies of the electron and the positron, and the direction of the positron are sampled by the rejection method, the final energies of the produced particles are then transformed to the laboratory frame, and the final energy of the proton is obtained by requiring energy conservation.

As for pion photoproduction, we have calculated the mean inelasticity for pair production, $\kappa_e(E)$, by averaging the fractional energy lost per interaction over a large number of simulated interactions of protons of energy $E$. We have then used $\kappa_e(E)$ and $\lambda_e(E)$ to calculate the loss rate for pair production in the same way as we did for pion photoproduction (see Equation 16).

## 3.4 Inverse Compton Losses

Consider a particle of mass $m$, charge $Ze$, and energy $E$ traversing the soft photon field in the central region of the AGN. Inverse Compton interactions will be in the "Klein-Nishina regime" if the energy of the incident photon in the particle's rest frame $\varepsilon' \simeq \gamma \varepsilon$ is greater than the particle's rest mass energy. Thus, for energies $E > E_{\text{crit}} = m^2 c^4 / \langle \varepsilon \rangle$, where $\langle \varepsilon \rangle$ is the mean photon energy, the Klein-Nishina cross section must be used. For spectrum (a) in particular, many of the inverse Compton interactions will take place in the Klein-Nishina energy regime, and so the energy loss rate due to inverse Compton scattering is given by (see e.g. Blumenthal and Gould [52])

$$\frac{dE}{dt} = \int_0^\infty \int_\varepsilon^{\varepsilon_1^{\max}} (\varepsilon_1 - \varepsilon) \frac{dN}{dt d\varepsilon_1} d\varepsilon_1 d\varepsilon, \qquad (20)$$

where $\varepsilon_1$ is the final energy of the photon measured in the laboratory frame, and $\varepsilon_1^{\max} = 4\gamma\varepsilon E/(mc^2 + 4\varepsilon\gamma)$ is given by the kinematics of the interaction. Here,

$$\frac{dN}{dt d\varepsilon_1} = \frac{2\pi r_0^2 cn(\varepsilon)}{\gamma^2 \varepsilon}\left[2q\ln q + (1+2q)(1-q) + \frac{(\Gamma q)^2}{2(1+\Gamma q)}(1-q)\right],$$

$$q = \frac{\varepsilon_1}{\Gamma(E - \varepsilon_1)},$$

$$\Gamma = \frac{4\varepsilon\gamma}{mc^2},$$

and $r_0$ is the classical radius of the particle.

## 3.5 Synchrotron Losses

Synchroton energy losses in the vicinity of the shock will be in the "Thomson regime" (by analogy with inverse Compton scattering) if the synchrotron photon's energy in the particle's rest frame is much less than rest mass energy of the particle, i.e.

$$\gamma heB/2\pi m \ll mc^2. \tag{21}$$

For the present model we obtain a critical energy below which this criterion is fulfilled:

$$E_{\rm crit} \simeq 4.1 \times 10^{-18} Z^{-1} (m/m_e)^3 x_1^{7/4} L_C^{1/2} Q^{1/2} \quad {\rm GeV}. \tag{22}$$

As we shall see in Section 3.7 where we calculate the maximum proton energy, for this dependence of $E_{\rm crit}$ on $L_C$ synchrotron losses are always in the Thomson regime in the central region of AGN. The energy loss rate is then given by the usual electron synchrotron formula (e.g.[53]) except that for protons the Thomson cross section of the electron is replaced by that of the proton. For the present model we obtain

$$\frac{dE}{dt} \simeq 5.2 \times 10^{36} Q^{-1} x_1^{-7/2} L_C^{-1} \left(\frac{mc^2}{\rm GeV}\right)^{-4} \left(\frac{E}{\rm GeV}\right)^2 \quad {\rm GeV\,s^{-1}}. \tag{23}$$

## 3.6 Other Loss Processes

In addition to the loss processes described above, particles in the central region may be advected into the black hole, or may diffuse out of the central region. If there is a relativistic wind in the AGN, then particles may also be advected away from the central region (e.g. ref. [38]) although we do not consider this possibility here.

If particles inside the central region are advected onto the black hole with a speed $u(r < R) \simeq u_2 = u_1/4$ then the time scale for advection from radius, $R$, is

$$t_{\rm adv} = R/u_2 \simeq 3.9 \times 10^{-43} x_1^{5/2} L_C \quad {\rm s}.$$

However, particles will be diffusing as they are advected, and during this time they will have typically diffused a distance, $r_{\text{diff}} = \sqrt{2Dt_{\text{adv}}}$, and so only a fraction of order $f \simeq r_S^2/r_{\text{diff}}^2$ will actually be advected into the black hole. Thus, for energies greater than

$$E_{\text{adv}} \simeq 1.8 \times 10^{-12} b^{-1} Z Q^{-1/2} L_C^{1/2} x_1^{-9/4} \qquad \text{GeV} \qquad (24)$$

the time scale for advection is effectively

$$\frac{t_{\text{adv}}}{f} \simeq 2.2 \times 10^{-31} b Z^{-1} Q L_C^{1/2} x_1^{19/4} (E/\text{GeV}) \qquad \text{s.} \qquad (25)$$

Particles will diffuse out of the central region of AGN on a time scale

$$t_{\text{diff}} \simeq \frac{R^2}{2D} \simeq 7.1 \times 10^{-55} b^{-1} Z Q^{-1/2} L_C^{3/2} x_1^{9/4} (E/\text{GeV})^{-1} \qquad \text{s.} \qquad (26)$$

However, once the mean path length for interaction with the irregular component of the magnetic field becomes comparable with the source size, diffusion is no longer appropriate. Thus, for energies greater than

$$E \simeq 7.2 \times 10^{-12} b^{-1} Z Q^{-1/2} x_1^{1/4} L_C^{1/2} \quad \text{GeV} \qquad (27)$$

the time scale for escape from the central region is given by the light crossing time of the source,

$$t_{\text{esc}} = R/c \simeq 9.8 \times 10^{-44} x_1^2 L_C \quad \text{s.} \qquad (28)$$

## 3.7 The Maximum Particle Energy in AGN

From Figures 2 and 3 we can see that the processes which dominate the loss rate of protons in AGN as the proton energy increases are: advection into the black hole, proton-proton interactions, proton-photon interactions and synchrotron radiation. To calculate the maximum proton energy achievable in the central region of AGN, $E_{\text{max}}$, we find the energy at which the acceleration rate equals the total loss rate. In Figure 5 we show how $E_{\text{max}}$ varies with the continuum luminosity. Results are given for both spectrum (a) and spectrum (b), and for shock radii corresponding to $10 \leq x_1 \leq 100$.

# 4 Neutron Escape from the Central Region of AGN

Neutrons are not confined by the magnetic fields within AGN and so they may escape from the central region. However, they are subject to $np$ and $n\gamma$ collisions. We have calculated the optical depth of AGN due to these loss processes to find the fraction of the neutrons able to escape to large radii.

We assume that the cross section for $np$ interactions is the same as for $pp$ interactions, and hence in the central region, where the number density is $n_2 \simeq 4n_1 = 4\rho_1/m_p$, the optical depth corresponding to distance $R$ is $\tau_{np}^{\rm in} \simeq 0.29 x_1^{-1/2} Q^{-1}$. Outside of the central region the matter density falls off as $\rho(r) = \rho_1 (r/R)^{-3/2}$ and the optical depth from radius $r = R$ to $r = \infty$ is $\tau_{np}^{\rm out} \simeq 0.14 x_1^{-1/2} Q^{-1}$. Thus, for $10 \leq x_1 \leq 100$, and the assumed acceleration efficiency, AGN are optically thin for $np$ interactions which may therefore be neglected in the present model.

We assume that the total cross section for neutron-initiated pion photoproduction is approximately the same as for proton-initiated pion photoproduction. Inside of the central region where the photon field is assumed to be isotropic, the optical depth for crossing distance $R$ is given simply by $\tau_{n\gamma}^{\rm in}(E) = R/\lambda_\pi(E)$. However, outside of the central region the situation is quite different. Here photons come primarily from the central region of the AGN and hence the photon field is highly anisotropic. In this case, the optical depth is given by

$$\tau_{n\gamma}^{\rm out}(E) = \int_R^\infty \frac{dr}{\lambda_\pi(E, r)}, \quad (29)$$

where $\lambda_\pi(E, r)$ is given by Equation 15 except that the minimum square of the CM frame energy is also dependent on radius and is given by

$$s_{\min} = m_n^2 c^4 + 2\varepsilon \left( E - pc\sqrt{1 - R^2/r^2} \right) \quad (30)$$

where $p$ is the neutron's momentum. This results from considering an isotropic radiation field but only integrating over interactions in which the photons originate in the central region. We show in Figure 6 the optical depths which result, for both spectrum (a) and spectrum (b), divided by the compactness parameter of the AGN which is usually defined as

$$l = \frac{L_C \sigma_T}{4\pi m_e c^3 R}. \quad (31)$$

In the present model, $l \simeq 730 x_1^{-2}$, and for $l \leq 1$ (i.e. $x_1 \geq 27$) all neutrons with energy less than $\sim 10^6$ GeV can escape from the central region without interaction. However, above $\sim 10^6$ GeV the probability of escape is sensitive to which continuum spectrum is present, and also on the compactness parameter.

## 5 The Monte Carlo Calculation

To obtain the expected neutrino and cosmic ray spectra from AGN, we first calculate the spectrum of neutrinos and cosmic rays produced per proton injected into the accelerator, and then scale the resulting spectra by the rate at which protons are injected. We consider separately particles produced during acceleration, and those produced after acceleration, as discussed below.

## 5.1 Particles Produced During Acceleration

We adopt a simple picture of shock acceleration in which we visualize the accelerator as a leaky-box into which particles of energy $E_0$ are injected and accelerated at a constant rate $a = dE/dt$, given by Equation 12. The time scale for escape from the leaky box is equal to the acceleration time scale, $t_{\rm esc} = t_{\rm acc} = E/a$, and so the probability of reaching an energy $E$ without escaping is $P_{\rm surv}(E) = E_0/E$. Such a model produces an $E^{-2}$ spectrum at energies above $E_0$ and may be easily adapted to calculations in which other processes take place in addition to acceleration, and are simulated by the Monte Carlo method.

To calculate the spectrum of particles produced during acceleration per injected particle, we use a method of weights. If a particle is injected with energy $E_0$, then the "weight" of the particle by the time it reaches an energy $E_1$ is set to the probability of it not having escaped, $W_1 = E_0/E_1$. If the particle interacts, and has energy $E_1'$ after the interaction, etc., then the weight of the particle at the $n$th interaction is given by

$$W_n = \frac{E_0}{E_1}\frac{E_1'}{E_2}\cdots\frac{E_{n-1}'}{E_n}. \tag{32}$$

For the maximum energies we are interested in ($10^6 \leq E_{\rm max} \leq 10^{10}$ GeV for protons) we can see from Figures 2 and 3 that pair production, pion photoproduction and synchrotron radiation are the most important loss processes.

For the case of extremely high acceleration rates synchrotron radiation limits the maximum proton energy. The synchrotron photons are typically in the 10 GeV energy range for $B \sim 10^3$ Gauss and $E \sim 10^9$ GeV and will contribute, along with electrons from pair production and electrons and photons from pion decay, to the electromagnetic cascade we assume gives rise to the AGN continuum spectrum. In this case, apart from synchrotron photons we assume the spectrum of particles leaving the accelerator is just the $E^{-2}$ proton spectrum extending up to $E_{\rm max}$.

For lower acceleration rates, for which $E_{\rm max}$ is determined by proton-photon collisions, we adopt the following procedure. We inject a particle with energy $E_0$ into the accelerator and after $(n-1)$ interactions (assuming it has survived catastrophic losses) it will have energy $E_{n-1}'$ and weight $W_{n-1}$. We sample the $n$th path length for interaction as follows. A dimensionless path length is sampled for both pion photoproduction and pair production using the sampling rule $\tau = -\ln\xi$, where $\xi$ is a random deviate distributed uniformly on [0,1]. The energy of the particle varies as a function of the distance travelled, $x$, and hence a pair of coupled equations must to solved to obtain the actual path length before the interaction. The rate of change of energy with distance is $dE/dx = a/c$, and hence $E = E_{n-1}' + ax/c$. Similarly, the rate at which the actual path length changes with the dimensionless path length is $dx/d\tau = \lambda(E)$, where $\lambda(E)$ is the mean path length for the interaction. We obtain the path length at which a particular interaction might take place by solving this equation numerically using the 4$^{\rm th}$ order Runge-Kutta method. The competing interaction with the shortest path length is then assumed to be the one that occurs.

Before the interaction, the proton has energy $E_n$ and the interaction is modelled as described in Section 3. The energies of the produced particles are binned in energy with the appropriate weight, in this case $W_n$. The 'fraction' of the particle which does not interact, in this case $(1 - W_n)$, is assumed to escape from the accelerator and have an $E^{-2}$ energy distribution over the energy range $E'_{n-1}$ to $E_n$. This contribution is then added to the escaping proton spectrum. The process is repeated until the particle suffers a catastrophic loss in the form of a pion photoproduction reaction in which a neutron is produced.

In this way we have calculated the spectra of protons, neutrons, pions (neutral and charged) and electrons (including positrons) produced during acceleration. Results for $E_{\max} = 10^8$ GeV, $x_1 = 30$, and spectrum (a) are shown in Figure 7.

## 5.2 Particles Produced After Acceleration

We adopt quite a different approach for calculating the spectrum of neutrinos produced by the protons and neutrons which escape from the accelerator. These particles will initiate a cascade in the matter and radiation field. Here, instead of following a single particle with a given weight, we calculate the energy spectrum of particles on interacting during the entire cascade, and get the total spectrum of neutrinos produced by convolving this with pre-calculated particle yields obtained with a Monte Carlo program. The problem naturally separates into finding the spectrum of neutrinos produced within the central region and finding the spectrum of neutrinos produced outside of the central region by neutrons which have escaped from the central region. We consider first particles produced within the central region.

Particles which have escaped from the accelerator lose energy in an electromagnetic–hadronic cascade through the matter, magnetic and radiation fields present. In this cascade we treat pair production as a continuous energy loss process along with synchrotron losses. The cascade will continue until the protons lose all of their kinetic energy or they are catastrophically lost, either by being advected into the black hole or undergoing a charge exchange interaction.

Consider the spectrum of protons escaping from the accelerator. For the $i^{\text{th}}$ energy bin centered on energy $E_i$ with bin width $\Delta E_i$, if the total effective energy loss rate is given by $dE/dt(E_i)$ then, on average, the time taken for a proton to "cross" the energy bin is $\Delta t = \Delta E_i/(dE/dt(E_i))$. Hence, if energy loss interactions occur at a rate $c/\lambda(E_i)$, where $\lambda(E_i)$ is the total mean path length for interaction, then the average number of interactions which occur in the energy bin is $n_i c \Delta t/\lambda(E_i)$, where $n_i$ is the number of protons which have cascaded down to the $i^{\text{th}}$ energy bin. In $n_i \Delta t c/\lambda_{p\gamma \to pX}(E_i)$ of these interactions, a proton remains in the final state, and the spectrum of protons produced in such interactions is simply added back to the total spectrum of protons. Similarly, $n_i \Delta t c/\lambda_{p\gamma \to nX}(E_i)$ protons are catastrophically lost. The remaining protons, i.e. those that cross the energy bin without interacting, are simply added to the $(i-1)^{\text{th}}$ energy bin.

We start with the highest energy bin of the accelerated proton spectrum and using the method outlined above calculate the spectrum of interaction energies which occur as the protons cascade down in energy. The spectrum of particles produced during the cascade can then be calculated by convolving the spectrum of interaction energies with pre-calculated particle yields. The particle yields give the probability of producing a particle of a given type in, say, the $j^{\text{th}}$ energy bin when a proton in the $i^{\text{th}}$ bin interacts with a proton or a photon. These yields have been calculated using a Monte Carlo program. The resulting particle spectra for $E_{\max} = 10^8$ GeV are shown in Figure 8 for spectrum (a). We have compared the results from this numerical treatment with those of our full Monte Carlo treatment and found excellent agreement. The use of the numerical approach as distinct from the full Monte Carlo method speeds up the calculation considerably.

As discussed in Section 4, ultra-relativistic neutrons can escape from the central region of the AGN where they decay, on average, after travelling a distance $\sim 2.9 \times 10^{13}(E/\text{GeV})$ cm. The resulting protons again initiate a hadronic cascade. However, in this case the interactions are primarily with protons, as the matter density falls off with radius as $r^{-3/2}$ whereas the radiation density falls off as $r^{-2}$ outside of the central region. Hence, in this subsequent cascade, we only consider nucleon-nucleon interactions and, for the purpose of this calculation, we do not distinguish between neutrons and protons. Again we calculate the spectrum of interactions throughout the cascade, except that in this case all of the particles in a given energy bin are assumed to interact via $pp$ or $np$ interactions. The spectrum of interactions is again convolved with the particle yields, and the resulting pion spectra are shown in Figure 9 for $E_{\max} = 10^8$ GeV under the assumption that no protons from neutron decay escape from the galaxy without interacting. As we will see in Section 5.4 below, this latter approximation breaks down above $10^5$ GeV, but its use here is adequate for calculating the neutrino spectrum below $10^5$ GeV.

## 5.3 The Neutrino Spectrum per Proton Injected into the Accelerator

In AGN the matter and radiation densities are sufficiently low such that all pions and muons produced decay before interacting. The two-body neutral pion decays are modelled exactly. To model the decay of the muons we neglect polarization effects and use the distributions of Zatsepin and Kuz'min [54]. The resulting spectra of muon and electron neutrinos (including antineutrinos) per particle injected into the accelerator are shown separately in Figure 10 for $E_{\max} = 10^6, 10^7, \ldots, 10^{10}$ GeV, $x_1 = 30$ and spectrum (a).

## 5.4 The Cosmic Ray Spectrum per Proton Injected into the Accelerator

As discussed above, neutrons may escape from the central region of the AGN and decay. The resulting protons will then diffuse in the magnetic field which is tied to the accreting plasma and some may eventually escape from the galaxy. The density of the accreting plasma falls off as $n(r) \propto r^{-3/2}$ while the temperature of the plasma falls off as $T(r) \propto r^{-1}$ [55]. Hence, if we assume that the pressure in magnetic turbulence tracks the plasma pressure then $D(r) \propto r^{5/2}$, giving

$$D(E,r) = 1.3 \times 10^{70} b Z^{-1} x_1^{-13/4} L_C^{-2} Q^{1/2} (E/\text{GeV})(r/\text{cm})^{5/2} \qquad \text{cm}^2\,\text{s}^{-1}, \qquad (33)$$

using Equation 10 for $D(E,R)$. The time scale on which protons are trapped in the accreting plasma is then approximately given by $t_{\text{esc}}(E,r) \simeq r^2/2D(E,r)$.

Protons will also be subject to proton-proton and proton-photon interactions and the time scales for these processes are calculated as described in Section 3, but with $\rho(r) \propto r^{-3/2}$ and $U_{\text{rad}}(r) \propto r^{-2}$. The probability of surviving the $pp$ and $p\gamma$ interactions and escaping from the active galaxy is

$$P_{\text{surv}} \simeq \frac{t_{\text{esc}}^{-1}}{t_{\text{esc}}^{-1} + t_{pp}^{-1} + t_{p\gamma}^{-1}}, \qquad (34)$$

where all of the time scales are calculated at a radius $r_{\text{dec}} \simeq 2.9 \times 10^{13}(E/\text{GeV})$ cm, the mean distance travelled before an ultrarelativistic neutron of energy $E$ decays.

We show in Figure 11 (solid lines) the spectrum of neutrons produced as a result of injecting one proton into the accelerator for the case where $b = 10$, $x_1 = 30$ and spectrum (a) for various luminosities. We use the probability of neutrons escaping from the central region without undergoing $p\gamma$ collisions, together with the probability of protons from neutron decay surviving $pp$ and $p\gamma$ collisions to obtain the spectrum of protons escaping from the enhanced density region of the accretion flow. We expect these protons will then readily escape from the host galaxy as extragalactic cosmic rays. We have added to Figure 11 (dashed lines) the escaping cosmic ray spectrum per injected proton.

# 6 The Intensity of Neutrinos and Cosmic Rays from AGN

The energy flux of neutrinos or cosmic rays (assuming a negligible intergalactic magnetic field) from a source at redshift $z$ is given by

$$E\frac{dF}{dE} = \frac{(1+z)}{4\pi d_L^2}\frac{dL}{dE'}, \qquad (35)$$

where $dL/dE'$ is the differential luminosity of neutrinos or cosmic rays produced by the AGN, $E = E'/(1+z)$ and $d_L$ is the luminosity distance. To calculate the differential spectrum of neutrinos or cosmic rays from a given AGN, we assume that the energy which goes into $\gamma$-rays and electrons ultimately produces the observed IR to hard X-ray continuum emission. Hence, the rate at which protons are injected into the accelerator may be approximated by

$$\dot{N} \simeq \frac{L_C}{W_{e\gamma}}, \qquad (36)$$

where $W_{e\gamma}$ is the energy which goes into electrons and $\gamma$-rays per proton injected into the accelerator. The differential spectrum is then obtained by multiplying the neutrino or cosmic ray spectrum per proton injected into the accelerator by $\dot{N}$.

To obtain the total contributions to the diffuse neutrino background and to the observed cosmic ray spectrum at Earth from the central regions of AGN, we need to integrate over the contributions from all epochs and over all AGN continuum luminosities, $L_C$. The luminosity function of AGN, i.e. number density per unit luminosity, is not available for the total AGN continuum luminosity. Instead, we use a luminosity function for the X-ray luminosity, $L_X$. The intensity of neutrinos or cosmic rays from all AGN is then given by

$$\frac{dI}{dE} = \frac{1}{4\pi} \frac{c}{H_0} E^{-1} \int dL_X \int_0^{z_{\max}} dz \frac{g(z)}{f(z)(1+z)^{5/2}} \rho_0\left(\frac{L_X}{f(z)}\right) \frac{dL}{dE}\{(1+z)E, kL_X\}, \quad (37)$$

for the Einstein-De Sitter model, where $\rho_0(L_X)$ is the local X-ray luminosity function of AGN, $g(z)$ and $f(z)$ describe its density and luminosity evolution as a function of redshift, and $H_0$ is the Hubble constant. Here, $k = L_C/L_X$ is the ratio of the AGN continuum luminosity to the X-ray luminosity. For both adopted AGN continuum spectra, $k \simeq 21$ for the 2 – 10 keV band. For the luminosity function, we have used the pure luminosity evolution model fits (models A to D) obtained by Morisawa *et al.* [56], and the pure luminosity evolution model fit obtained by Maccacaro *et al.* [57], and in each case, we take the values of $H_0$, $q_0$ and $z_{\max}$ given in these papers appropriate to the model used. The resulting intensities of neutrinos and cosmic rays are shown in Figures 12 and 13, respectively. In each case, we show by the shaded band the range between the lowest and highest intensities predicted for the range of luminosity functions considered, the range of $x_1$ adopted, as well as the variation between results for spectrum (a) and spectrum (b).

## 6.1 The Neutrino Intensity

In Figure 12 we show our result together with the background due to atmospheric neutrinos [58]. With our present assumption, magnetic pressure equals ram pressure of upstream accreting plasma, the maximum energies are about a factor of two higher than our previous results [24] in which we assumed the magnetic energy density to equal the radiation energy density at radius $r = R$. The original results of Stecker *et al.* [20] were too high by a factor of $\sim 17$ due to an error in the luminosity function they used (see Szabo and Protheroe [24] for details; see also refs. [25, 59, 38, 60]). We compare our results with those of Stecker *et al.* [26], Sikora and Begelman [38], Biermann [60], and results for blazars due to Stecker [26] and Nellen *et al.* [61]. In the present work the X-ray luminosity functions used give an AGN contribution to the observed X-ray background of 100% (Morisawa *et al.* [56]) or 40% (Maccacaro *et al.* [57]). In the work Stecker *et al.* [20], Stecker *et al.* [25], and Sikora and Begelman [38] the contributions are 100%, 20% and 20%, respectively. These differences are partly responsible for discrepancies in overall normalization. In addition, the discrepancy at high energy with Stecker *et al.* arises because of their use of $B = 10^3$ gauss for all AGN (and not including synchrotron loss) which gives rise to their higher maximum energies. At low energies the discrepancy is due to their assumption that protons will escape from the central region if the straight-line optical depth to $p\gamma$ collisions is less than unity, whereas we assume they are magnetically trapped and all interact. At low energies, Sikora and Begelman assume protons are advected onto the black hole or away from the AGN at speed $c$ rather than $u_2$ (see Section 3.7). At high energies, their result differs from ours, partly because of their approximation that all AGN are located at one redshift. The result of Biermann is based on scaling from the observed X-ray background using energetics arguments. In view of these very different approaches to the problem, the results are consistent in as much as the origin of the differences is well understood.

## 6.2 The Cosmic Ray Intensity

Cosmic rays with energies below about $10^{10}$ GeV are usually thought to be of galactic origin, those below a rigidity (momentum/charge) of $\sim 10^5$ GV being due to acceleration at supernova shocks. Between $10^6$ GeV and $10^7$ GeV, the 'knee' in the cosmic ray spectrum, the spectrum steepens from $\sim E^{-2.7}$ to $\sim E^{-3}$. At around $10^{19}$ eV the spectrum flattens, possibly due to an extragalactic component [62] of cosmic rays accelerated at shocks in jets of active galactic nuclei or lobes of radio galaxies [63, 64, 65, 66]. Between $10^6$ GeV and $10^{10}$ GeV the origin is unknown, although it has been suggested to be due to acceleration by supernova shocks in expanding magnetized stellar wind cavities [67] or reacceleration of the low-energy galactic population [68, 69].

In Figure 13 we compare results from the present work with a summary of the observations [70]. From this figure we see that in the region of the "knee", at $\sim 10^7$ GeV, one would expect a substantial fraction of the observed cosmic rays to originate in AGN. Note that our predictions shown in Figure 13 have not been normalized in any way to the cosmic ray data and are determined solely by the accretion/shock acceleration model used and by the observed X-ray luminosity function of AGN. We find it quite remarkable that the predicted contribution from AGN to the cosmic ray spectrum is of the same order of magnitude as the observed intensity in the region of the knee.

Extragalactic cosmic rays are subject to galactic modulation, but for reasonable galactic wind parameters this is unimportant for rigidities above $\sim 10^5$ GV (W.-H. Ip, personal communication). One uncertainty in our prediction stems from our complete ignorance of intergalactic propagation. Diffusion coefficients at $10^{16}$ eV in the range $D \sim 10^{31} - 10^{37}$ cm$^2$ s$^{-1}$ have been mentioned in the literature [1, 71, 72, 73, 74]. AGN appear to have been present over much of the age of the universe, $t_U$, based on the observation of AGN at large redshift. In this time, cosmic rays may diffuse a distance $r_{\max} \simeq (2Dt_U)^{1/2}$. Based on the observed 2 − 10 keV luminosity function [56], the number density of AGN is about $4 \times 10^{-3}$ Mpc$^{-3}$. For $D > 3 \times 10^{34}$ cm$^2$ s$^{-1}$ the number of AGN within $r_{\max}$ is greater than 500 and the result of Figure 13 will apply.

For $D < 3 \times 10^{34}$ cm$^2$ s$^{-1}$ the situation is more complicated. Even though at any epoch the number of objects displaying AGN activity within $r_{\max}$ may be small, AGN can only be active for $\sim 1\%$ of $t_U$, and so the total number of AGN may be a factor of $\sim 100$ higher if one includes those now dormant. These dormant AGN would, however, have contributed to the observed cosmic rays. So, we could argue that our results may apply, even for diffusion coefficients as low as $10^{33}$ cm$^2$ s$^{-1}$. Nevertheless, individual nearby sources active during recent epochs may play an important role in shaping the observed spectrum if $D < 3 \times 10^{34}$ cm$^2$ s$^{-1}$ and one should really take account of the time dependence of these sources [73]. We may, however, get some indication of the intensity due to nearby sources for the case of such low diffusion coefficients by considering a typical source at a distance corresponding to the average spacing between AGN. For a source at distance $d$ continually active for a time larger than the diffusion time from the source, the intensity at Earth, neglecting cosmological effects, is

$$\frac{\mathrm{d}I_{\mathrm{CR}}}{\mathrm{d}E} \simeq \frac{c}{(4\pi)^2 Dd} E^{-1} \frac{\mathrm{d}L_{\mathrm{CR}}}{\mathrm{d}E}. \tag{38}$$

We have added to Figure 13 (dotted lines) results for a typical AGN at $d = 10$ Mpc for $D \sim 10^{33} - 3 \times 10^{34}$ cm$^2$ s$^{-1}$. For this, we have taken an X-ray luminosity of $L_X = 10^{43}$ erg s$^{-1}$, $b = 10$, $x_1 = 30$ and spectrum (a). As can be seen, the contribution from nearby sources can more than make up for the lack of contributions from distant sources. For this case, the anisotropy, $\delta$, outside our galaxy at $10^7$ GeV, would be much less than that of the single source, $\delta \ll 3D/cd \sim (0.3\% - 10\%)$, because of contributions from different sources in different directions at different epochs. In any case, the anisotropy would be reduced drastically by diffusion within our galaxy and is not in conflict with observations; we point out here that above $10^6$ GeV there appears to be no convincing evidence that any real cosmic ray anisotropy has yet been observed [75].

## 7 Discussion and Conclusions

It turns out that our results are rather insensitive to the way we estimate the magnetic field at the shock. If we had assumed the magnetic energy density at the shock to be equal to the radiation energy density then we would have obtained only a slightly lower acceleration rate, a factor of $\sim 3$ lower for $x_1 = 10$ and $\sim 6$ lower for $x_1 = 100$, giving maximum energies lower by factors of $\sim 1.7$ and $\sim 2.5$ respectively. Perhaps the largest uncertainty in our calculations arises due to the uncertainty of about a factor of two in the value of $L_C/L_X$ adopted. For example, if $L_C/L_X$ were increased by a factor of two then the main effect on our results would be to increase the predicted intensities by approximately a factor of two at all energies.

We would like to emphasize that our calculation is made using a simple model of AGN which may approximate the conditions in some AGN central engines in which accretion onto a supermassive black hole takes place. Our calculations do not apply to AGN models such as 'spinars' or 'magnetoids' although particle acceleration has been invoked in such models and would give rise to neutrino and cosmic ray production. Also, some AGN models do not permit proton acceleration up to the energies discussed in the present work, if at all. Clearly, the high energy neutrino background and cosmic ray proton intensity calculated for such models would be much less than given in the present work which should therefore be considered as an estimate of the maximum contribution from AGN central engines, as distinct from AGN jets.

In conclusion, we have calculated the spectrum of neutrinos produced in AGN per low energy proton injected into the accelerator. The pioneering calculation of Stecker *et al.* [20] stimulated much interest because it predicted fluxes observable with DUMAND and AMANDA. Unfortunately, their original fluxes needed to be revised down by a factor of $\sim 17$ making detection marginal. However, our inclusion of interactions of protons down to lower energies than assumed by Stecker *et al.*, leads to higher intensities at 1 TeV and we predict detectable neutrino fluxes. Already data on horizontal air showers are able to rule out the the high intensities initially predicted by Stecker *et al.* at PeV energies [76], and data on horizontal underground muons from the Frejus experiment are able to rule out some models at TeV energies [77]. The next few years should prove very interesting with the DUMAND, AMANDA, NESTOR and Baikal projects being commissioned.

We also conclude that AGN may be an important source of cosmic rays in the region of the knee. The enhancement which appears to be present at the knee may be due to this extragalactic component. In addition to protons, heavier nuclei will also be accelerated. However, no heavy nuclei will escape from the central region as they will be broken up in interactions with photons in the central region. Thus, any extragalactic component in the region of the knee will be 100% protons. If the present model is correct, then one would expect to observe an enhancement in the relative abundance of protons in the cosmic rays at $\sim 10^7$ GeV. At present the data in this energy range are indirect, being based on air shower observations and observations of muons deep underground. The most recent air shower data are from the Mt. Norikura array (Saito *et al.* [78]) and suggest that at $10^7$ GeV, the proton component is only about 10% of the total cosmic ray intensity. Data below $10^5$ GeV from the JACEE Collaboration [79] and just above $10^8$ GeV from the Fly's Eye experiment [80] appear also show a low proton component, although the cosmic rays appear to become systematically lighter with increasing energy above $3 \times 10^8$ GeV [81]. We have added these data to Figure 13, and note that if the air shower data [78] are confirmed, they will place severe limits on the acceleration rate in AGN, or on the fraction of AGN in which acceleration takes place. The present air shower data do not rule out the present model, but would require an acceleration rate corresponding to $b > 100$, or acceleration taking place in fewer than 10% of all AGN. However, first indications from a recent study of the cosmic ray composition using the MACRO detector at Gran Sasso [82] favor a light composition at energies up to several thousand TeV which, if extrapolated, would be consistent with even our highest acceleration rate (corresponding to $b = 1$) in all AGN. Future experiments may well be able to measure the composition at the knee more directly and put stricter limits on our model.

# Acknowledgements

R.J.P. is grateful to W.-H. Ip, T. Stanev and G. Zank for helpful discussions. This work was supported by a grant from the Australian Research Council. We thank P.A. Johnson for helpful comments on the manuscript.

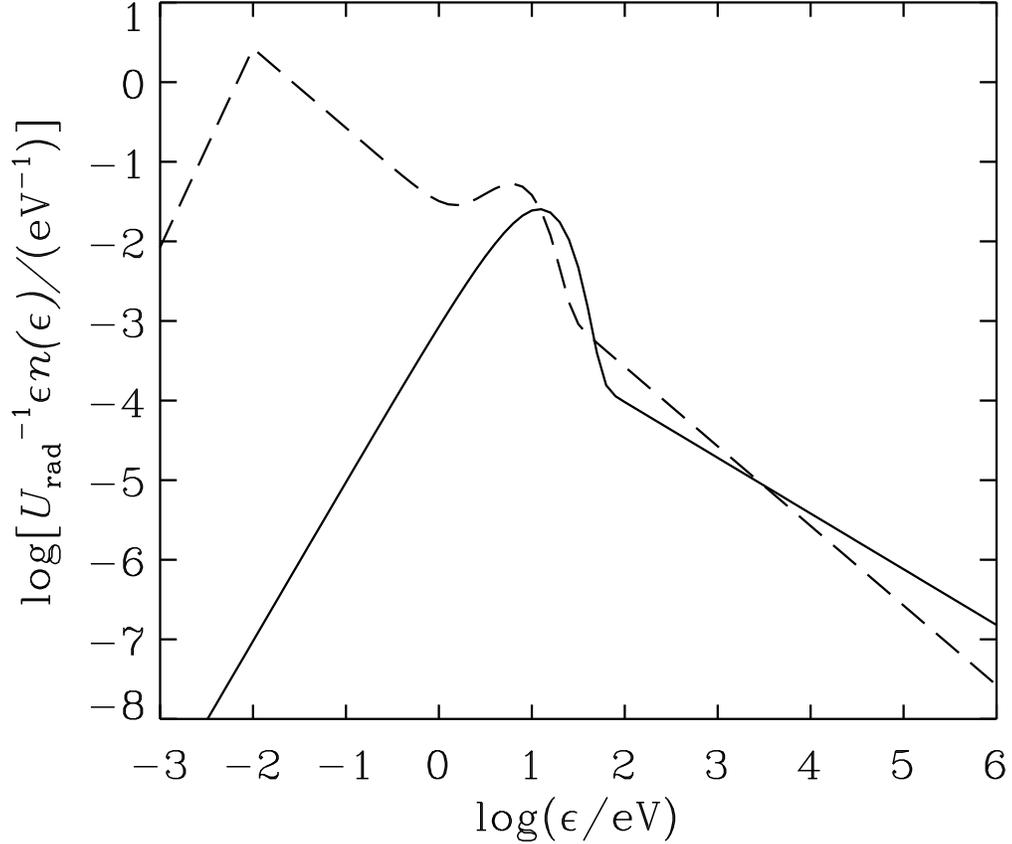

Figure 1: A plot of the two "generic" photon spectra used to model the radiation present in the central region of AGN. $n(\varepsilon)$ is the differential photon number density. The solid curve represents spectrum (a): an $\varepsilon^{-1.7}$ photon spectrum for $10 \leq \varepsilon \leq 10^6$ eV plus a thermal spectrum with temperature $T = 5 \times 10^4$ K. The dashed curve represents spectrum (b): an $\varepsilon^{-2}$ photon spectrum for $10^{-2} \leq \varepsilon \leq 10^6$ eV plus a thermal spectrum with temperature $T = 2.6 \times 10^4$ K; for $\varepsilon < 10^{-2}$ eV, we use an $\varepsilon^{3/2}$ photon spectrum. In each case, we assume equal energy density in the power-law and thermal components.

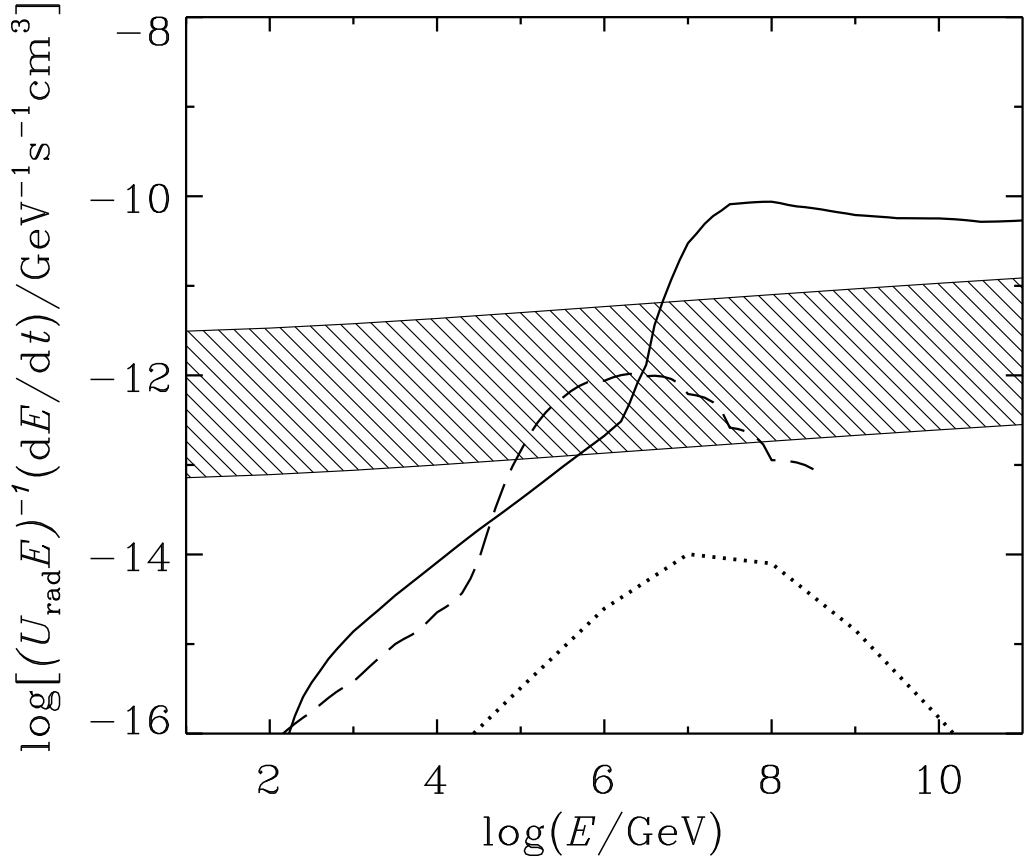

Figure 2: The loss rates for protons in the central region of AGN scaled by the radiation energy density. The curves give the loss rates for pion photoproduction (solid curve), pair production (dashed curve), inverse Compton scattering (dotted curve) and proton-proton interactions (hatched band). Results are shown for spectrum (a) and $L_C^{1/2}/b = 10^{21}$ (erg s$^{-1}$)$^{1/2}$; the hatched band corresponds to the range $10 \leq x_1 \leq 100$.

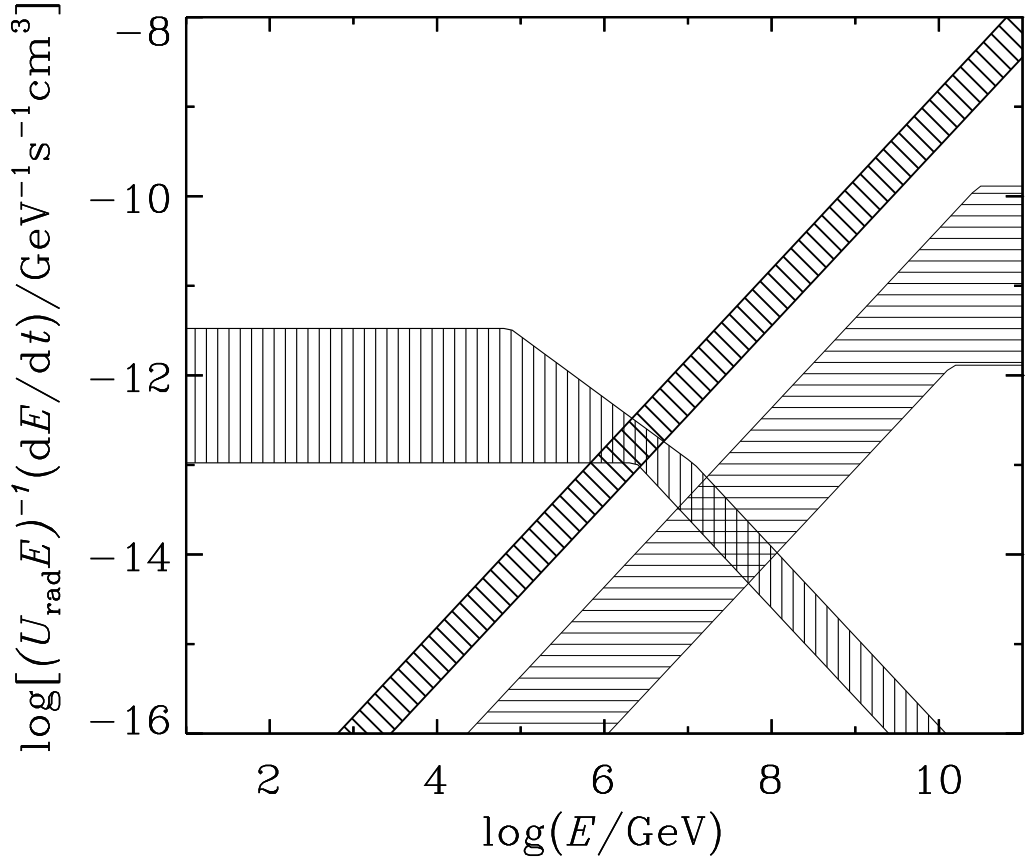

Figure 3: The loss rates for protons in the central region of AGN scaled by the radiation energy density. The hatched bands give the loss rates for synchrotron radiation (oblique hatching), advection onto the black hole (vertical hatching) and diffusion away from the central region (horizontal hatching). Results are shown for spectrum (a) and $L_C^{1/2}/b = 10^{21}$ (erg s$^{-1}$)$^{1/2}$; each of the hatched bands correspond to the range $10 \leq x_1 \leq 100$.

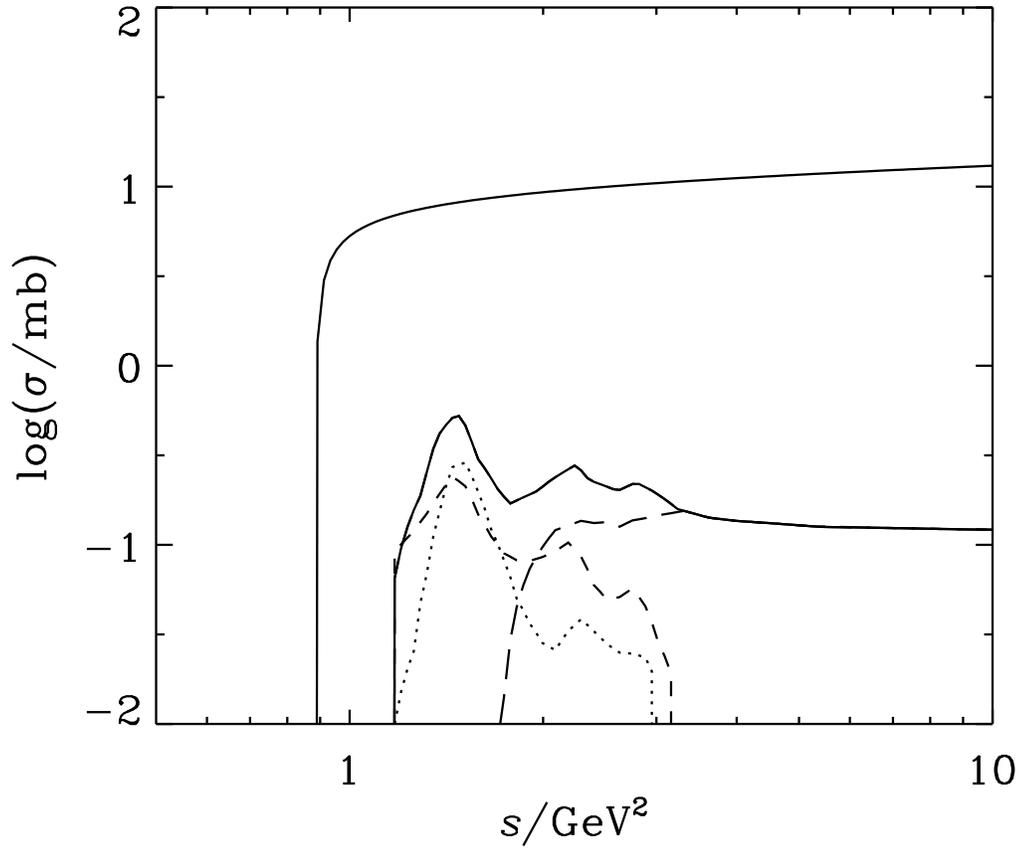

Figure 4: The total cross section for pion photoproduction (lower full curve) as a function of $s$, and the contributions from $\gamma p \to \pi^0 p$ (dotted line), $\gamma p \to \pi^+ n$ (short dashed line), and $\gamma p \to$ multiple pion production (long dashed line). Also shown is the total cross section for pair production (upper full curve).

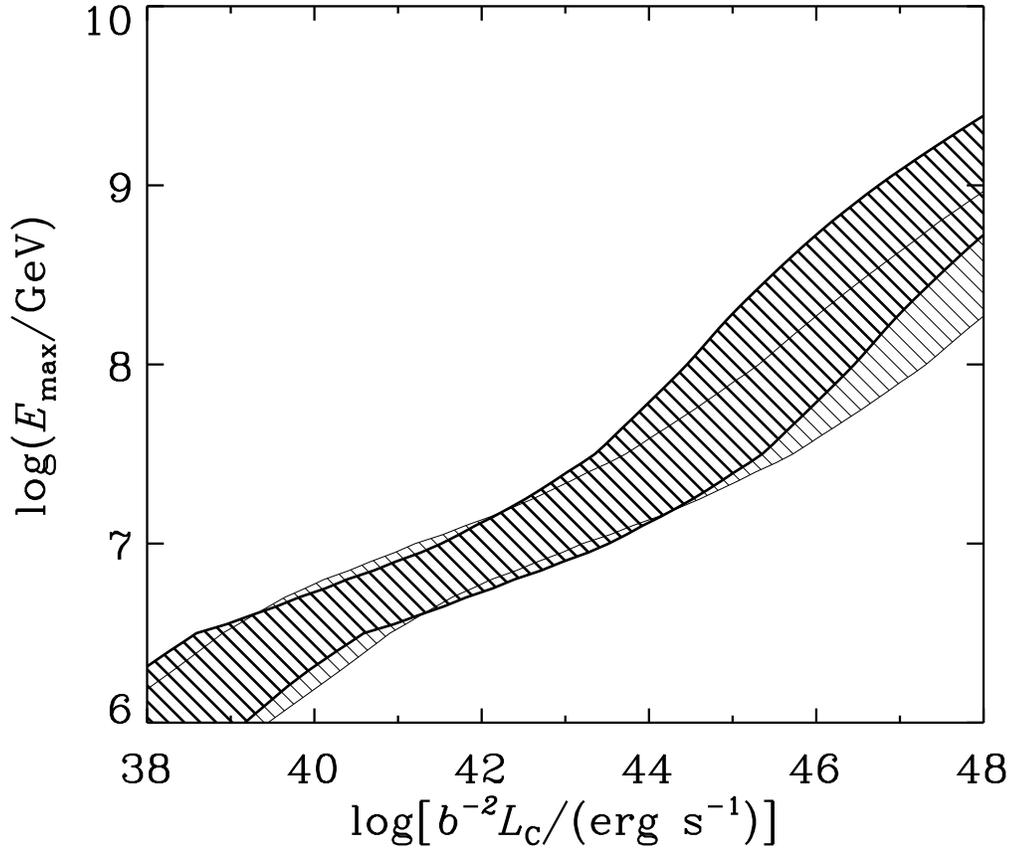

Figure 5: The maximum proton energy as a function of the AGN continuum luminosity, $L_C$ for spectrum (a) (thickly hatched region between thick full curves) and for spectrum (b) (thinly hatched region between thin full curves).

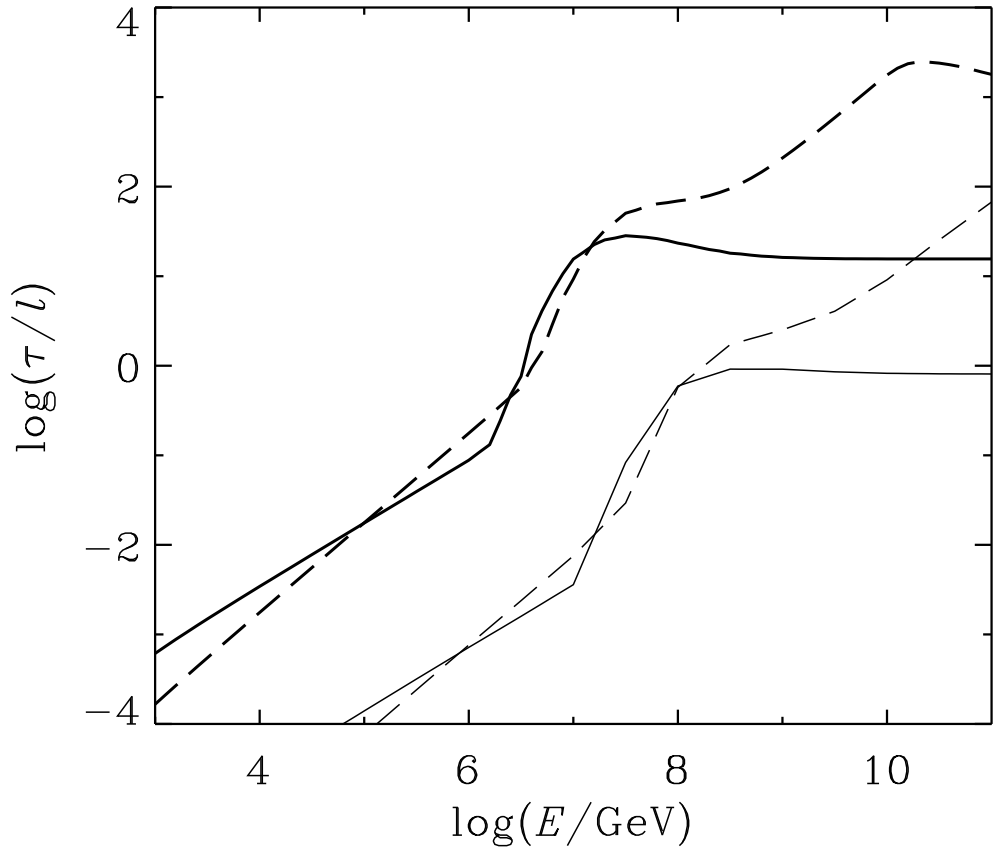

Figure 6: The optical depth (divided by the compactness parameter) inside the central region (thicker curves) and outside the central region (thinner curves) due to neutron initiated pion photoproduction. Results are shown for both spetrum (a) (full curves) and spectrum (b) (dashed curves).

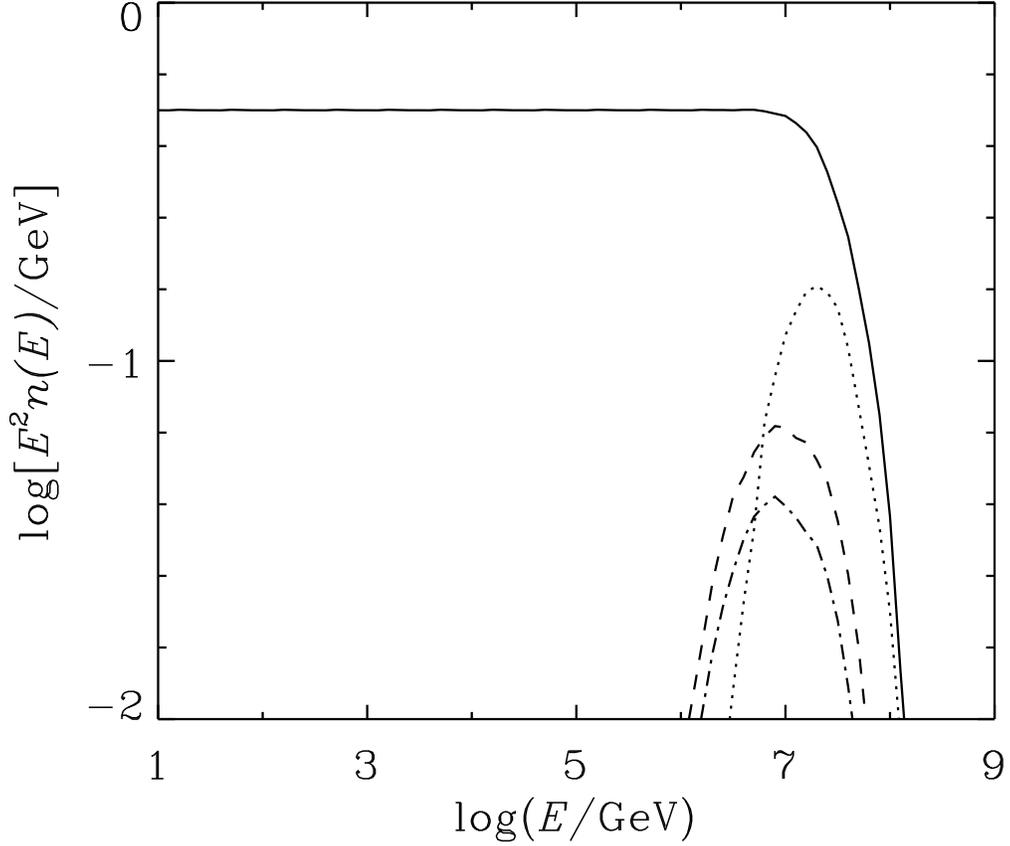

Figure 7: The spectrum of particles produced during acceleration (multiplied by $E^2$) per proton injected into the accelerator: protons (full curve), neutrons (dotted curve), charged pions (dashed curve) and neutral pions. Results are shown for AGN continuum spectrum (a), $x_1 = 30$ and $E_{\mathrm{max}} = 10^8$ GeV (i.e. acceleration rate equals total effective energy loss rate at $E = 10^8$ GeV; note this does not preclude a small fraction of particles reaching slightly higher energies).

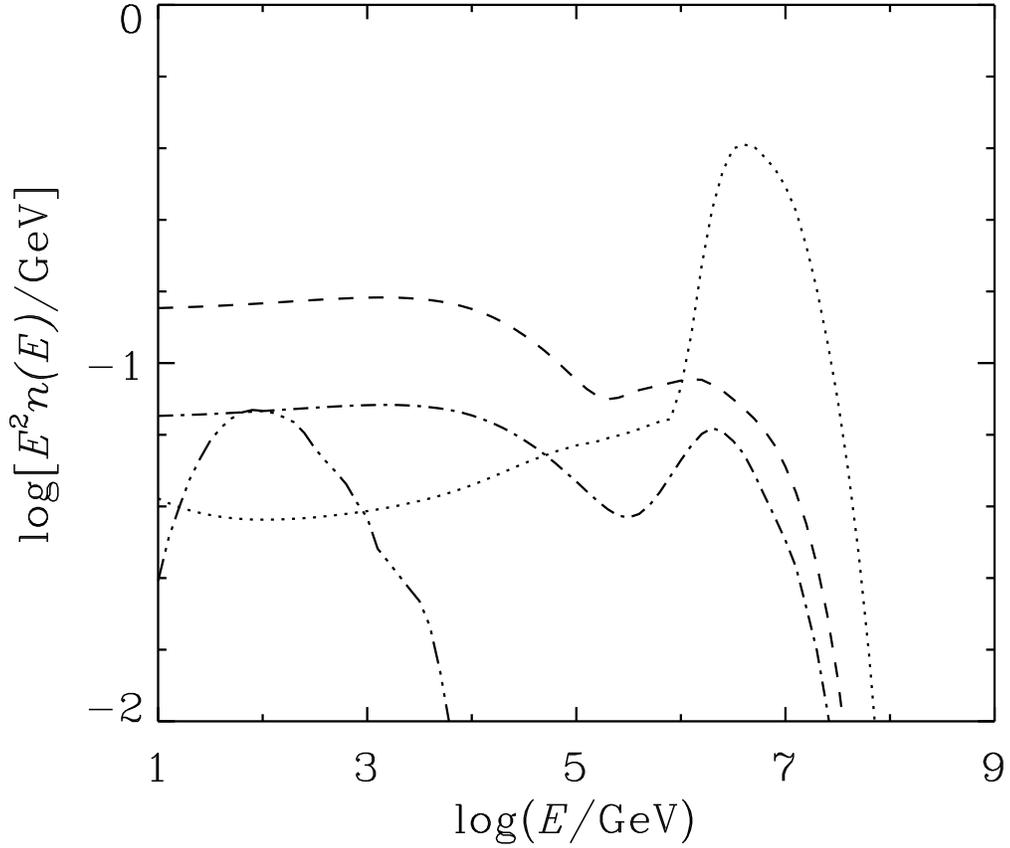

Figure 8: The spectrum of particles produced within the central region (multiplied by $E^2$) per proton injected into the accelerator: neutrons (dotted curve), charged pions (dashed curve), neutral pions (dot dashed curve), electrons (e$^\pm$) from pair production (dot dot dot dashed curve). Results are shown for spectrum (a), $E_{\mathrm{max}} = 10^8$ GeV, and $x_1 = 30$.

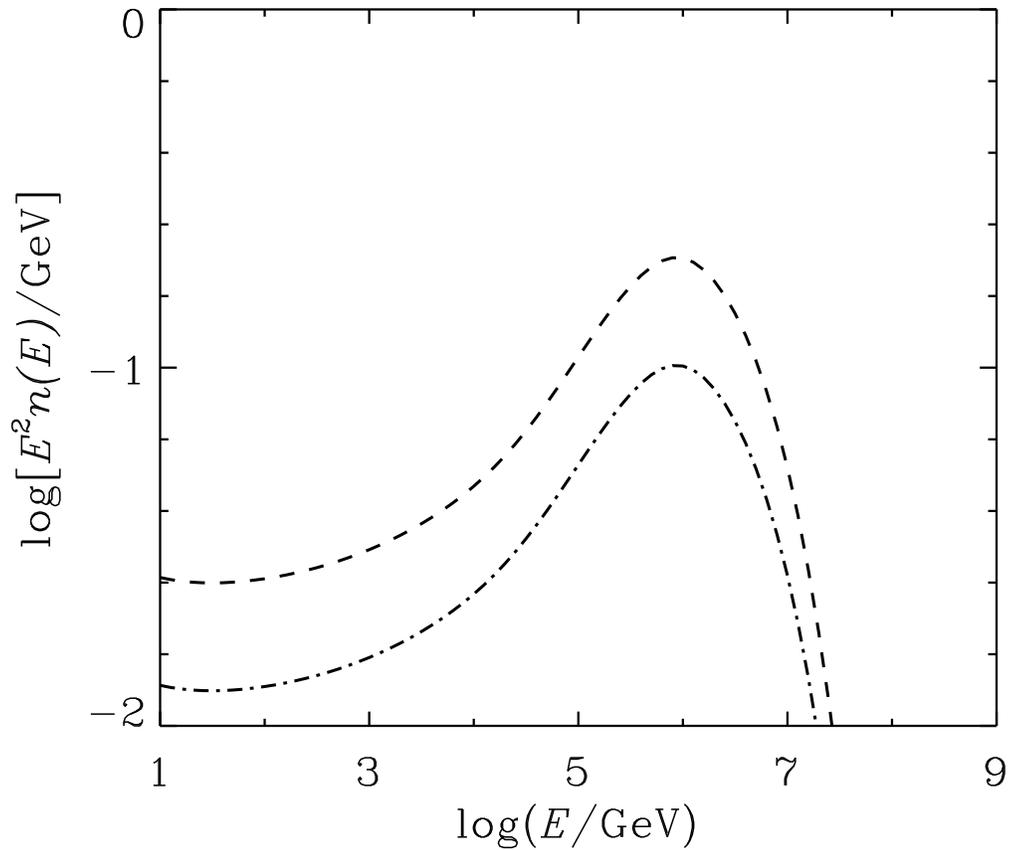

Figure 9: The spectrum of particles produced by neutrons which have escaped from the central region (multiplied by $E^2$) per proton injected into the accelerator: charged pions (dashed curve) and neutral pions (dot dashed curve). Results are shown for spectrum (a), $E_{\max} = 10^8$ GeV, and $x_1 = 30$.

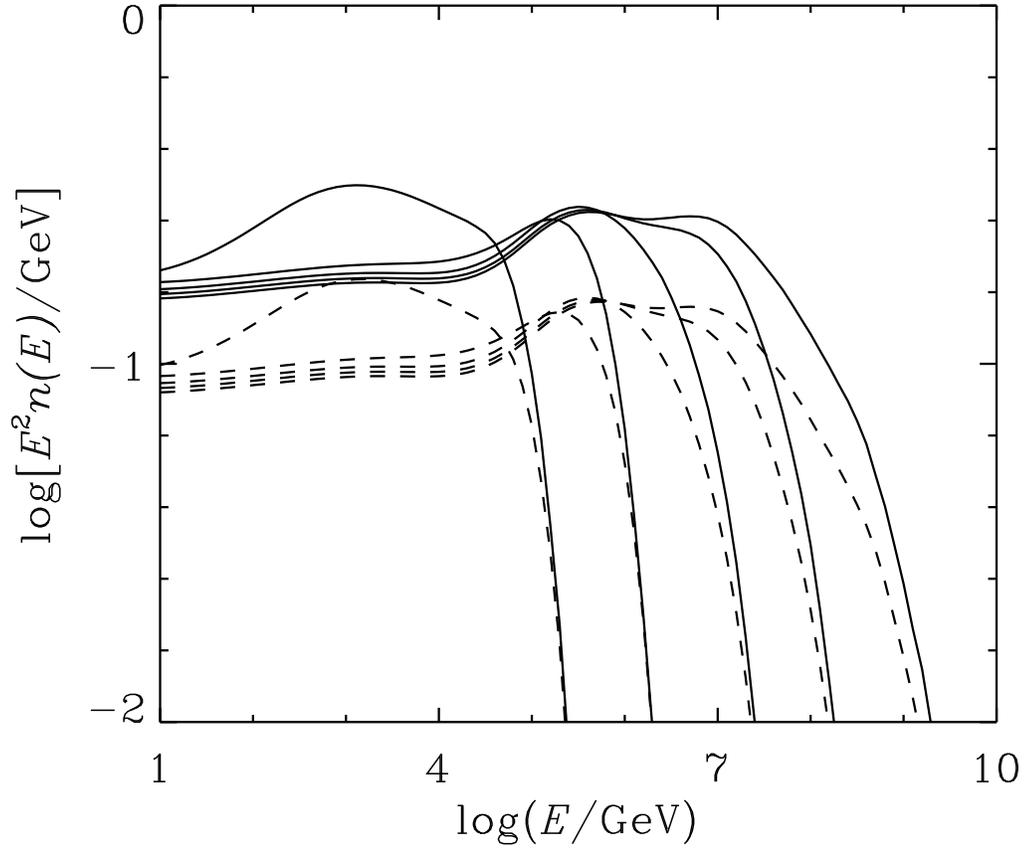

Figure 10: The spectra of muon neutrinos (full curves) and electron neutrinos (dashed curves) (multiplied by $E^2$) produced per proton injected into the accelerator. Results are shown for spectrum (a), $x_1 = 30$ and $E_{\mathrm{max}} = 10^6$ (left-most curves), $10^7$,..., $10^{10}$ GeV.

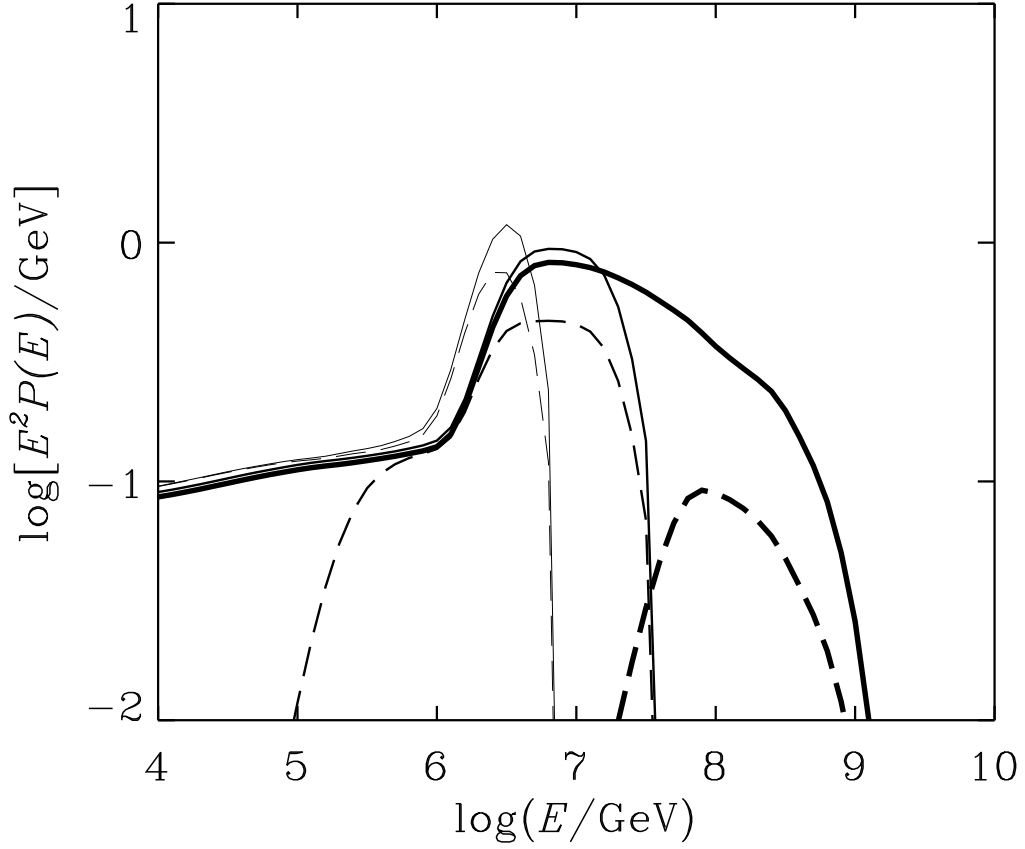

Figure 11: Spectra of neutrons produced (full curves) and cosmic ray protons escaping from an AGN (dashed curves) per low energy proton injected into the accelerator. Results are multiplied by $E^2$, and are given for $b = 10$, $x_1 = 30$, spectrum (a) and $L_X = 10^{42}$ (leftmost curves), $10^{45}$, and $10^{48}$ erg s$^{-1}$. In each case, $P(E)dE$ gives the number in the range $E$ to $(E + dE)$ per injected proton. The reduction at high energies is due to interactions of neutrons with photons during escape from the central region, while the reduction at low energies is due to interactions of protons from neutron decay with protons in the accreting plasma.

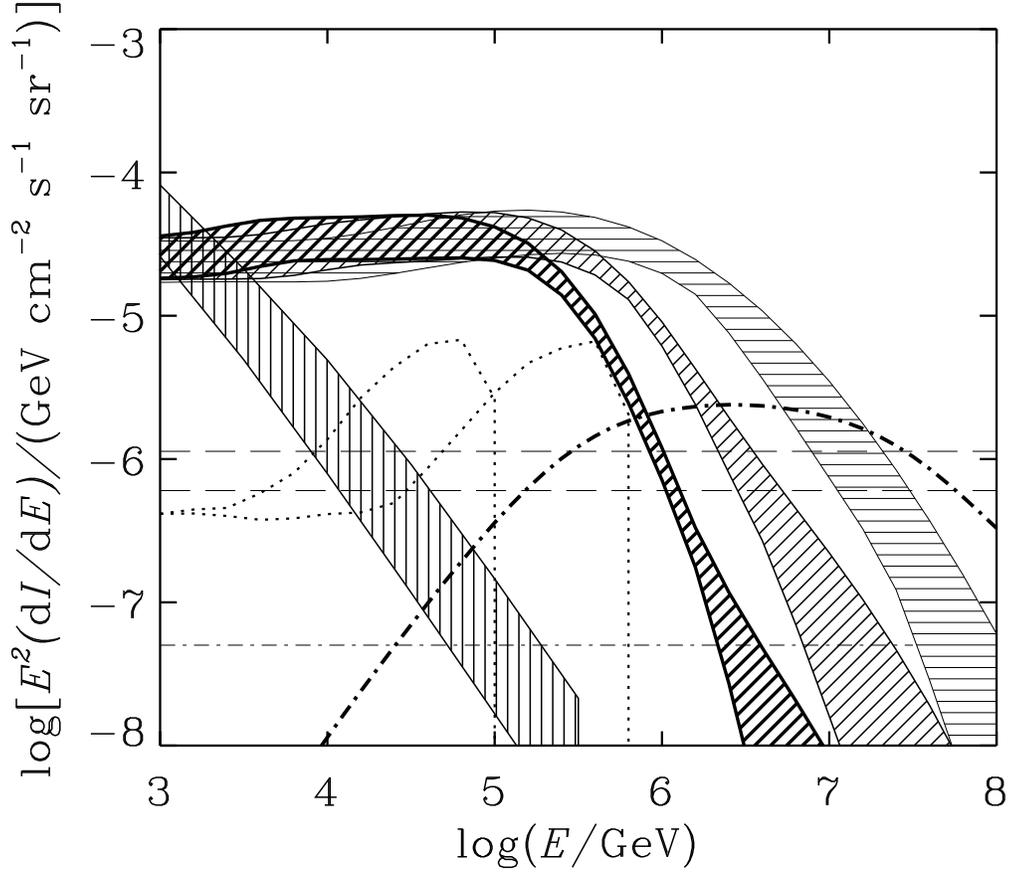

Figure 12: The expected diffuse $\nu_\mu + \bar{\nu}_\mu$ intensity at Earth. The hatched bands show the spread in results obtained using both spectrum (a) and spectrum (b), and models (a) to (e) of Morisawa *et al.* [56] and the model of Maccacaro *et al.* [57] for the luminosity function. Results are shown for $b=1$ (horizontal hatching), $b=10$ (thin oblique hatching) and $b=100$ (thick oblique hatching). An integration over a flat distribution in $\log x_1$ has been made for $10 < x_1 < 100$. Also shown: Stecker *et al.* [25,26] (chain curve), Sikora and Begelman [37] (dotted curves) for sources at $z = 0$ and 5, Biermann [60] (lower dashed line), and blazar contributions calculated by Stecker [26] (chain line), and Nellen et al. [61] (upper dashed line). The atmospheric neutrino intensity [58] is shown by the vertical hatched band: the upper curve corresponds to zenith angle $\theta = 90°$ and the lower curve corresponds to $\theta = 0°$.

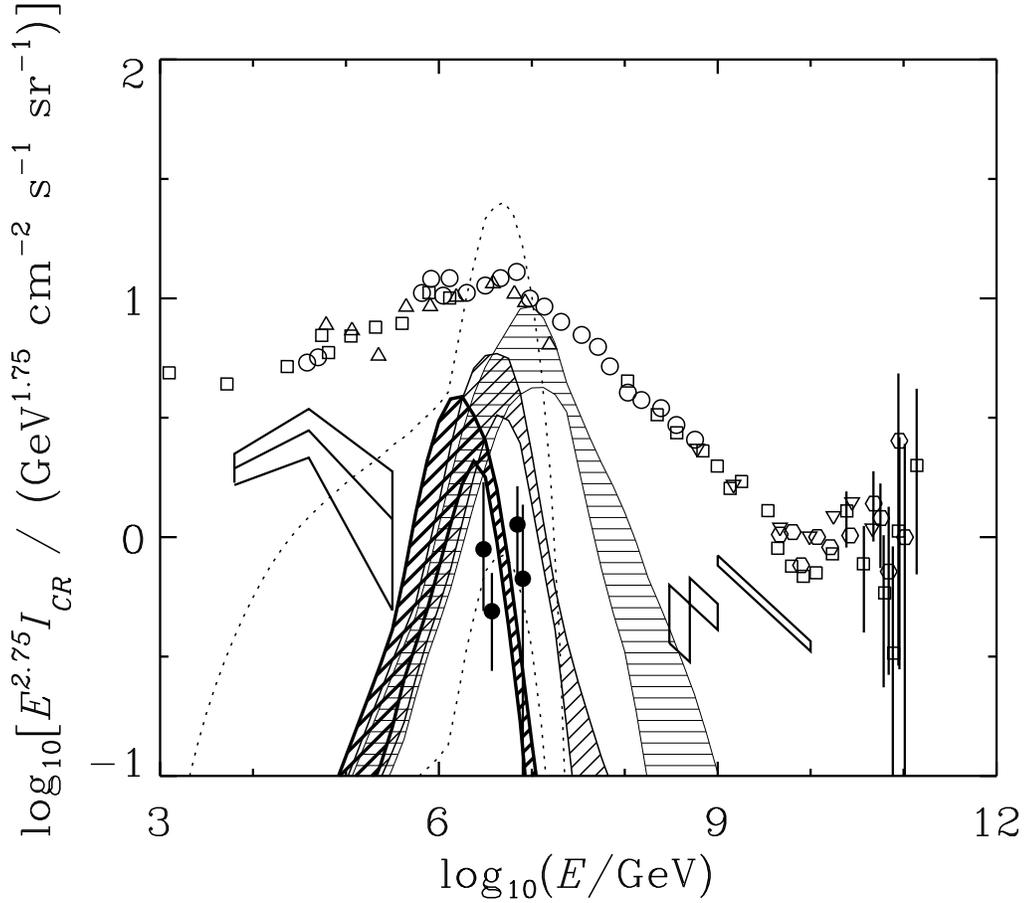

Figure 13: The possible contribution of cosmic rays accelerated in AGN to the observed spectrum. The meaning of the hatched bands is the same as in Figure 12. An integration over a flat distribution in $\log x_1$ has been made for $10 < x_1 < 100$. Observations of the total cosmic ray intensity (open symbols) are from the survey by Stanev [70]. The proton intensity for $6 \times 10^3$ GeV to $3 \times 10^5$ GeV as measured by the JACEE Collaboration [79] (broken power-law with error box), and estimates of the proton intensity from the Mt. Norikura air shower array [78] (filled circles) and the Fly's Eye experiment [80] (rectangles above $10^8$ GeV) are shown. We also show our results (dotted lines) for a single steady source at $d = 10$ Mpc (see text). For this we use $D \sim 10^{33}$ cm$^2$ s$^{-1}$ (upper curve) and $3 \times 10^{34}$ cm$^2$ s$^{-1}$ (lower curve). Source parameters are $L_X = 10^{43}$ erg s$^{-1}$, $b = 10$, $x_1 = 30$, and spectrum (a).

37